\documentclass[12pt]{article}
\pdfoutput=1

\usepackage{amsmath}
\usepackage{amssymb}
\usepackage{graphicx}

\def\fv{\mathbf{5}}
\def\bfv{\mathbf{\bar{5}}}

\def\for{\mathbf{ 45}}
\def\bfor{\mathbf{\bar{45}}}
\def\se{\mathbf{ 70}}
\def\bse{\mathbf{\bar{70}}}
\def\tf{\mathbf{24}}

\def\bSigma{\mathbf{\Sigma}}

\newcommand{\AddrLNF}{
  {\it INFN, Laboratori Nazionali di Frascati,C.P. 13, I00044
    Frascati, Italy}}
\newcommand{\AddrIFT}{
  {\it 
IFT-UAM/CSIC,  
Nicolas Cabrera 15, C.U. Cantoblanco,  28049 Madrid, Spain
}}
\newcommand{\AddrUAM}{
  {\it 
Departamento de F\'{\i}sica Te\'orica,                                
                   Universidad Aut\'onoma de Madrid, } \\ 
 {\it                C.U. Cantoblanco, 28049 Madrid, Spain 
}}
\newcommand{\AddrUdeA}{
  {\it Instituto de F\'{\i}sica, Universidad de Antioquia,
    A.A.{\it{1226}}, Medell\'{i}n, Colombia}}

\title{
\rightline{
\vbox{
  \vspace{-2.5cm}
\hbox{\normalsize FTUAM-11-46\qquad IFT-UAM/CSIC-11-27}
  \vspace{2.0cm}
}}
Neutrino masses in $SU(5)\times U(1)_F$ with adjoint flavons}
\author{Enrico Nardi$^{1,2,3}$, Diego Restrepo$^4$, and Mauricio Velasquez$^4$\\
$^1$\AddrLNF\\
$^2$\AddrIFT\\
$^3$\AddrUAM\\
$^4$\AddrUdeA}
\date{}

\begin{document}
\maketitle

\begin{abstract}
  We present a $SU(5)\times U(1)_F$ supersymmetric model for neutrino
  masses and mixings that implements the seesaw mechanism by means of
  the heavy $SU(2)$ singlets and triplets states contained in three
  adjoints of $SU(5)$.  We discuss how Abelian $U(1)_F$ symmetries can
  naturally yield non-hierarchical light neutrinos even when the heavy
  states are strongly hierarchical, and how it can also ensure that
  $R$--parity arises as an exact accidental symmetry.  By assigning
  two flavons that break $U(1)_F$ to the adjoint representation of
  $SU(5)$ and assuming universality for all the fundamental couplings,
  the coefficients of the effective Yukawa and Majorana mass operators
  become calculable in terms of group theoretical quantities.  There
  is a single free parameter in the model, however, at leading order
  the structure of the light neutrinos mass matrix is determined in a
  parameter independent way.
\end{abstract}

\baselineskip 18pt

\section{Introduction}

The standard model (SM) is a very successful framework for describing
particle physics phenomena. However, it suffers from some serious
theoretical problem, among which: neutrinos are massless, the
conditions for baryogenesis are not fulfilled, and there is no
candidate for the dark matter (DM).  The first two problems can be
solved by extending the SM to include the seesaw mechanism for
neutrino masses~\cite{Minkowski:1977sc,Yanagida:1979as,Glashow,%
  GellMann:1980vs,Mohapatra:1980yp} that also opens the possibility of
baryogenesis via leptogenesis~\cite{Fukugita:1986hr,Davidson:2008bu},
while extending the SM to its supersymmetric version (SSM) can provide
a natural candidate for DM.  However, in contrast to the SM, the SSM
does not have accidental lepton ($L$) and baryon-number ($B$)
symmetries, and this can lead to major phenomenological problems, like
fast proton decay.  The standard solution to forbid all dangerous
operators is to impose a discrete symmetry, $R$--parity, and only in
the $R$-parity conserving SSM the lightest supersymmetric particle (LSP),
generally  the neutralino, is stable, and provides a good DM
candidate.

Similarly to the SM, also the SSM does not provide any explanation for
the strong hierarchy in the fermion Yukawa couplings.  One way to
explain the flavor puzzle and the suppression of the fermion masses
with respect to the electroweak breaking scale is to impose Abelian
flavor symmetries, that we generically denote as $U(1)_F$, that are
broken by SM-singlets commonly denoted as flavons. Besides the Yukawa
couplings, these symmetries can also suppress, but often not forbid
completely, the SSM $B$ and $L$ violating terms. Along these lines,
consistent models can be build in which small neutrino masses can be
accommodated (for a review see \cite{Dreiner:2003hw}). Due to the fact
that in these models $R$--parity is not an exact symmetry, the LSP can
decay, however, long lived LSP can also provide acceptable DM
candidates~\cite{Sierra:2009zq}.

When $R$ parity is not imposed, there is also the possibility that it
could arise as an accidental symmetry like it happens in the SM for
$B$ and $L$.  $R$-parity conservation can be for example enforced by
an extended gauge symmetry together with supersymmetry (that requires
a holomorphic superpotential) as in the model studied
in~\cite{Mira:1999fx}, or solely by the gauge symmetry thanks to a
suitable choice of the $U(1)_F$--charges, as in
ref.~\cite{Dreiner:2003yr}.  In this paper we focus on this second
possibility, and we implement it in the framework of a unified
$SU(5)\times U(1)_F$ model.  A virtue of the $U(1)_F$ gauge symmetry
of our model is that when the $U(1)_F$ charges are chosen
appropriately, $\Delta B\neq 0$ and $\Delta L=1$ operators are
forbidden at all orders.  However, $\Delta L=2$ operators
corresponding to Majorana masses for heavy neutral fermions of the
seesaw remain allowed, and thus the seesaw mechanism can be embedded
in the model.  More in detail, following~\cite{Dreiner:2003yr} we
chose the $F$-charges in such a way that operators with even
$R$--parity have an overall $F$-charge that is an integer multiple of
the charge of the $U(1)_F$ breaking scalar fields (that, without loss
of generality, we set equal to $\pm 1$).  In contrast, all the
$R$--parity breaking operators, that have an overall half-odd-integer
$F$--charge, are forbidden. Then, to allow for $\Delta L=2$ Majorana
masses while forbidding $\Delta L=1$ operators, it is sufficient to
chose the $F$--charges of the heavy seesaw neutral states as
half--odd-integers.

Differently from the SM case~\cite{Dreiner:2003yr}, in $SU(5)$ GUTs it
is rather difficult to implement this kind of horizontal symmetries,
because there is less freedom in choosing the $F$--charges (see for
example~\cite{Chen:2008tc}). However, if the flavons that break the
horizontal symmetry are assigned to the adjoint representation of
$SU(5)$~\cite{Aristizabal:2003zn,Duque:2008ah,Wang:2011ub}, charges
that were forbidden in the singlet flavon case become allowed, under
the assumption that certain representations for the Froggatt-Nielsen
(FN)~\cite{Froggatt:1978nt} messengers fields do not exist.  In
contrast to the non-unified $SU(3)\times SU(2)\times U(1)\times
U(1)_F$ model, where the singlet nature of the flavons is mandatory,
in $SU(5)\times U(1)_F$ assigning the flavons to the adjoint has the
additional bonus that non-trivial group theoretical coefficients
concur to determine the coefficients of the effective
operators~\cite{Aristizabal:2003zn,Duque:2008ah,Wang:2011ub}.  In this
case, under the additional assumption that at the fundamental level
all the Yukawa couplings obey to some principle of
universality~\cite{Duque:2008ah}, the order one coefficients that
determine quantitatively the structure of the mass matrices become
calculable. In this paper we will avoid all speculations concerning
the fundamental physics that might underlie such a universality
principle; we just take it as a convenient working hypothesis: turning
off the `noise' related to the usual incalculable order one Yukawa
couplings allows to put in clear the role played by the calculable
group theoretical coefficients that multiply all the relevant
effective operators.

\section{Theoretical framework}

\subsection{Same sign and both signs Abelian charges}
\label{sec:signs}

Sometimes symmetry considerations are sufficient to determine
univocally the structure of the low energy operators, however, other
times a detailed knowledge of the full high energy theory is needed.
Let us consider for example a $U(1)_F$ symmetry and assume that all
the heavy and light states have charges of the same sign, say
positive. Then a single spurion $\epsilon_{-1}$ with a negative unit
charge is involved in the construction of all $U(1)_F$ (formally)
invariant operators. Let us consider the ${\rm dim}=5$ seesaw operator
$ {\cal L}_{D5} \sim - \frac{g_{\alpha\beta} }{2\,M}\left(\bar
  \ell_\alpha H\right) \left( H^T\ell^c_\beta\right) $, where
$\ell_\alpha$ are the lepton doublets and $H$ is the Higgs field, that
for simplicity we take neutral under the Abelian symmetry $F(H)=0$.
Since the only spurion useful to construct (formally) invariant
operators is $\epsilon_{-1}$, one can easily convince himself that the
structure of $g_{\alpha\beta}$, and thus the structure of the light neutrino
mass matrix, is univocally determined by the $F$ charges of the
light leptons as: $g_{\alpha\beta} \sim \epsilon_{-1}^{
  F(\ell_\alpha)+ F(\ell_\beta)}$, while the $F$-charges of whatever
heavy states of mass $\sim M$ are inducing the effective operator are
irrelevant.\footnote{It should be remarked that, contrary to what is
  sometimes stated, Abelian $U(1)_F$ symmetries allow to arrange very
  easily for non-hierarchical light neutrinos together with strongly
  hierarchical heavy neutrinos (as are often preferred in
  leptogenesis) by simply choosing $F(\ell_\alpha)=F(\ell)$ for all
  $\alpha$, and $ F(N_1)\gg F(N_2)\gg F(N_3)$.}  We can conclude that
in this case one does not need to consider the details of the high
energy theory, since the structure of the low energy effective
operators can be straightforwardly read off from the charges of the
light states.

However, if we allow for $U(1)_F$ charges of both signs, then both
symmetry breaking spurions $\epsilon_{-1}= \epsilon_{+1}= \epsilon$
are relevant. This implies that naive charge counting applied to the
low energy effective operators is unreliable, since basically a factor
$\epsilon^n$, as estimated in the low energy theory, could correspond
instead to $\epsilon^{n+m}_{+1} \cdot \epsilon^m_{-1} \sim
\epsilon^{n+2m}$.  Clearly the naive estimate can result in a
completely different (and wrong) structure with respect to the one
effectively generated by the high energy theory.  We illustrate this
with a simple example: let us take two lepton doublets with charges
$F(\ell_1)= -F(\ell_2)= +1$ and again $F(H)=0$.  The structure of the
light neutrino mass matrix read off from the lepton doublets charges
would be given by the low energy coefficient:
\begin{equation}                     
  \label{eq:mnu1}                                                 
   g_{\alpha\beta}  
\sim \begin{pmatrix}  
\epsilon^2 & 1 \\ 1 & \epsilon^2 \end{pmatrix} \,.
\end{equation}
This corresponds to a pair of quasi degenerate (pseudo-Dirac) light
neutrinos.

Now, let us assume that the fundamental high energy (seesaw) theory
has two right handed  neutrinos with charges $F(N_{1,2})=+1$.
For the heavy mass matrix $M_N$, its inverse, and for the Yukawa coupling
$Y_{\alpha i}\bar\ell_\alpha N_i$ we obtain:
\begin{align}
  \label{eq:MN}
M_N &\sim \epsilon^2               
\begin{pmatrix} 
1&1\\1&1
\end{pmatrix}\,,&
M_N^{-1} &\sim   \epsilon^{-2}
\begin{pmatrix}1&1\\ 1&1\end{pmatrix}\,,&
Y &\sim  
\begin{pmatrix}1&1\\ \epsilon^2&\epsilon^2\end{pmatrix}\,.
\end{align}
The resulting effective low energy coefficient  is: 
\begin{equation}                                                  
  \label{eq:mnu2}                                                 
g_{\alpha\beta}  \sim                                                    
Y M_N^{-1}Y^T  \sim  
 \epsilon^{-2}
\begin{pmatrix}
1 & \epsilon^2 \\
\epsilon^2 & \epsilon^4
\end{pmatrix}
\,,
\end{equation}
which (for $\epsilon\ll 1$) corresponds to very hierarchical and
mildly mixed light neutrinos, that is a completely different result
from the previous one.   

The model we are going to describe in this paper requires fermions
with charges of both signs, as well as a pair of positively and
negatively charged spurions. Therefore a detailed knowledge of the
high energy theory is mandatory, and accordingly we will explicitly
describe all its relevant aspects.

\subsection{Outline of the $SU(5)\times U(1)_F$ model}

We assume that at the fundamental level all the Yukawa couplings are
universal, and that all the heavy messengers states carrying $U(1)_F$
charges have the same mass, as it would happen if the masses are
generated by the vacuum expectation values (vev) of some singlet
scalar.  With these assumptions, the only free parameter of the model
is the ratio between the vacuum expectation value of the flavons and
the mass of the heavy vectorlike FN fields. This parameter is
responsible for the fermion mass hierarchy, and all the remaining
features of the mass spectrum are calculable in terms of group
theoretical coefficients.  More precisely, in our model the flavor
symmetry is broken by vevs of scalar fields $\langle
\Sigma_{\pm}\rangle$ in the $\mathbf{24}$--dimensional adjoint
representation of $SU(5)$, where the subscripts refer to the values
$\pm 1$ of the $U(1)_F$ charges that set the normalization for all the
other charges. The vevs
$\langle\Sigma_+\rangle=\langle\Sigma_-\rangle=V_a$ with
$V_a=V\cdot\operatorname{diag}(2,2,2,-3,-3)/\sqrt{60}$ are also
responsible for breaking the GUT symmetry down to the
electroweak--color gauge group. The size of the order parameters
breaking the flavor symmetry is then $\epsilon=V/M$ where $M$ is the
common mass of the heavy FN vectorlike fields. This symmetry breaking
scheme has two important consequences: power suppression in
$\epsilon$ appear with coefficients related to the different entries
in $V_a$, and the FN fields are not restricted to the $\mathbf{5}$,
$\overline{\mathbf{5}}$, or $\mathbf{10}$, $\overline{\mathbf{10}}$,
multiplets as is the case when the $U(1)_F$ breaking is triggered by
singlet flavons~\cite{Aristizabal:2003zn,Duque:2008ah}.

The model studied in \cite{Duque:2008ah} adopted this same scheme, and
yields a viable phenomenology, since it produces quark masses and
mixings and charged lepton masses that are in agreement with the data.
The $U(1)_F$ charge assignments of the model yield $U(1)_F$ mixed
anomalies, that are canceled trough the Green-Schwartz
mechanism~\cite{Green:1984sg}. The values of the charges are
determined only modulo an overall rescaling, that may be appropriately
chosen in order to forbid baryon and lepton number violating
couplings. However, with the choice of charges adopted
in~\cite{Duque:2008ah}, both $\Delta L=1$ and $\Delta L=2$ violating
operators were forbidden, and thus the seesaw mechanism could not be
embedded in the model. In order to avoid this unpleasant feature, in
this work we explore the possibility of forbidding just the $\Delta
L=1$ operators while allowing the $\Delta L=2$ seesaw operator for
neutrino masses.  We will show that by means of a suitable choice of
the $F$ charges, the seesaw mechanism can be implemented, and one can
obtain neutrino masses and mixings in agreement with oscillation data,
while $\Delta L=1$ and $\Delta B\neq 0$ (and thus $R$--parity
violating) operators are forbidden at all orders by virtue of the
$F$-charges.  Moreover, the scale of the heavy seesaw neutral fermions
remains fixed, and lies a few order of magnitude below the GUT scale,
and is of the right order to allow the generation of the baryon
asymmetry through leptogenesis.

\subsection{Charge assignments}

The  $F$ charges have to satisfy some specific
requirements in order to yield a viable phenomenology.  In the
following we denote for simplicity the various $F$ charges with the
same label denoting the corresponding $SU(5)$ multiplet.  To
allow a Higgsino $\mu$--term at tree level, we must require
\begin{align}
\label{eq:mu}
  \overline{\mathbf{5}}^{\phi_d}+\mathbf{5}_{\phi_u}=0\,,
\end{align}
where $ \overline{\mathbf{5}}^{\phi_d},\,\mathbf{5}_{\phi_u}$ denote
the $F$-charges of the 
chiral multiplets containing the $SU(2)$  Higgs doublets 
$\phi_d,\,\phi_u$. 
It is easy to see that with the constraint~(\ref{eq:mu}) the
overall charge of the Yukawa operators for the charged fermion masses
$ \mathbf{10}_{I} \mathbf{\bar{5}}_{J} \mathbf{\bar{5}}^{{\phi}_{d}}$
and $ \mathbf{10}_{I} \mathbf{10}_{J} \mathbf{{5}}_{{\phi}_{u}}$, that
are even under $R$--parity, are invariant under the charge
redefinitions~\cite{Duque:2008ah}:
\begin{align}
\label{eq:shift}
      \mathbf{\bar{5}}_{I}&\to \mathbf{\bar{5}}_{I} + a_n\\ \nonumber
    \mathbf{10}_{I} &\to  \mathbf{10}_{I} -\frac{a_n}{3}\\ \nonumber
    \mathbf{\bar{5}}^{{\phi}_{d}} &\to
    \mathbf{\bar{5}}^{{\phi}_{d}} - \frac{2a_n}{3}\\ \nonumber
    \mathbf{5}_{{\phi}_{u}}&\to \mathbf{5}_{{\phi}_{u}} +\frac{2a_n}{3}\,,
\end{align}
where $I=1,2,3$ is a generation index, and $a_n$ is an arbitrary
parameter that can be used to redefine the charges.  Assuming
$\mathbf{5}_{\phi_u}= 0$,
then the anomalous solution that
was chosen in ref.~\cite{Duque:2008ah} can be written as
\begin{align}
  \label{eq:1}
\mathbf{5}_{\phi_u}= \overline{\mathbf{5}}^{\phi_d} = &0\,, &&
\overline{\mathbf{5}}_I=2^I-7\,, &&\mathbf{10}_I=3-I\,.
\end{align}

Starting from a set of integer charges, and redefining 
this set by means of the shift 
eq.~(\ref{eq:shift}) with 
\begin{align}
  \label{eq:2}
  a_n=-\frac{3}{2}\left( \frac{2n}{5} +1\right)\,,
\end{align}
where $n$ is an integer, it is easy to see that the $R$--parity
violating operators $\mathbf{10}_{I} \mathbf{\bar{5}}_{J}
\mathbf{\bar{5}}_{K}$ and $\mathbf{\bar{5}}_{I}
\mathbf{{5}}_{{\phi}_{u}}$ have half--odd--integer charges, and hence
are forbidden at all orders by the $U(1)_F$ symmetry. 

To generate neutrino masses, we now introduce three heavy multiplets
$\mathbf{N}_I$ $(I=1,2,3$) with half--odd--integer $F$--charges, that
we assume corresponding to adjoint representations $\mathbf{24}$.  The
adjoint of $SU(5)$ contains two types of $SU(2)$ multiplets that can
induce at low energy the dimension five Weinberg
operator~\cite{Weinberg:1979sa}: one $SU(2)\times U(1)\times SU(3)$
singlet that allows to implement the usual type I seesaw, and one
$U(1)\times SU(3)$ singlet but $SU(2)$ triplet giving rise to a type
III seesaw~\cite{Bajc:2006ia,Bajc:2007zf,Biggio:2010me}. Contributions
from these two types of multiplets unavoidably come together, so that
by assigning `right handed neutrinos' to the $\mathbf{24}$ of $SU(5)$
one necessarily ends up with a type I+III seesaw.\footnote{We thank
  the referee for bringing this point to our attention.}
%
%
This slightly more complicated seesaw structure is not crucial for our
construction, but we still keep track of it for a matter of
consistency.

The half--odd--integer charges of the new states, after the charges of
the other fields have been shifted according to eqs.~(\ref{eq:shift})
and (\ref{eq:2}), can be parameterized as
\begin{align}
\label{eq:3}
N_I=\frac{2m_I + 1}{2}\,,
\end{align}
where $m_I$ are integers. The  {\it effective} superpotential terms
that give rise to the seesaw are
\begin{align}
\label{eq:4}
  W_{\text{seesaw}}=Y_\nu^{I J}\,\overline{\mathbf{5}}_{I}\,\mathbf{5}_{\phi_u}\,\mathbf{N}_{J}
+\tfrac{1}{2}M_R^{I J} \mathbf{N}_I \mathbf{N}_J\,.
\end{align}
The coefficient $Y_\nu^{I J}$ of the Dirac operator in
eq.~\eqref{eq:4} is determined by the following sums of $F$--charges:
\begin{align}
\overline{\mathbf{5}}_{I}+\mathbf{5}_{\phi_u}+\mathbf{N}_{J}
=&2^I-7+a_n+2a_n/3+N_J\nonumber\\
=&2^I-9+n+m_J\,.\nonumber
\end{align}
Explicitly:
\begin{align}
\label{eq:5}
  F(\overline{\mathbf{5}}_{I}\,\mathbf{5}_{\phi_u}\,\mathbf{N}_{J})=&
\begin{pmatrix}
    -7-n+m_1 &  -7-n+m_2 & -7-n+m_3 \\
   -5-n+m_1 &  -5-n+m_2 & -5-n+m_3 \\
   -1-n+m_1 &  -1-n+m_2 & -1-n+m_3 \\
\end{pmatrix}.
\end{align}
For the mass operator of  the adjoint neutrinos 
we have the following (integer) $F$--charges
\begin{align}
\label{eq:6}
   \mathbf{N}_{I}+\mathbf{N}_{J}=&1+m_I+m_J\,,\nonumber\\
  F(\mathbf{N}_I\,\mathbf{N}_J) =&\begin{pmatrix}
    1+ 2m_1   &1+ m_1+m_2 &1+m_1+m_3  \\
    1+ m_1+m_2 &1+ 2m_2  &1+m_2+m_3\\
    1+ m_1+m_3 &1+ m_2+m_3  &1+2m_3\\
   \end{pmatrix}.
\end{align}
The light neutrino
mass matrix is then obtained from the seesaw formula
\begin{align}
  \label{eq:7}
  M_\nu \approx& - {v^2}\sin^2\beta\,Y_\nu\,M_R^{-1}\,Y^T_{\nu}\,,
\end{align}
where $v=175\,$GeV, and it is left understood that in eq.~\eqref{eq:7}
the contributions of the $SU(2)$ singlets and triplets are both summed
up.
As is implied by the FN mechanism, the order of magnitude of the
entries in $Y_\nu$ and $M_R$ is determined by the corresponding values
of the sums of $F$ charges eqs.~(\ref{eq:5}) and
(\ref{eq:6}) as:
\begin{align}
  \label{eq:8}
  Y_\nu^{I J}&\sim\epsilon^{|\overline{\mathbf{5}}_{I}+\mathbf{5}_{\phi_u}+\mathbf{N}_{J} |}
\nonumber\\
M_R^{I J}&\sim M 
\cdot\epsilon^{|\mathbf{N}_{I}+\mathbf{N}_{J}|} = 
V \cdot\epsilon^{|\mathbf{N}_{I}+\mathbf{N}_{J}|-1}\,.
 \end{align}
 where in the second relation $M$ is the mass of the FN messengers
 fields and in the last equality we have used $M=\epsilon^{-1} V$.
 Note that since we have two flavon multiplets $\Sigma_{\pm}$ with
 opposite charges, the horizontal symmetry allows for operators with
 charges of both signs, and hence the exponents of the symmetry
 breaking parameter $\epsilon$ in eq.~\eqref{eq:8} must be given in
 terms of the absolute values of the sum of charges.  In FN models
 only the order of magnitude of the entries in eq.~\eqref{eq:8} are
 determined, and it is generally assumed that non-hierarchical order
 one coefficients multiply each entry.  However, in our model the
 assumption of universality for the fundamental Yukawa couplings has
 been made in order to avoid arbitrary $\mathcal{O}(1)$ numbers of
 unspecified origin.\footnote{This condition excludes the simple (and
   often used) charge assignments in which there are two zero
   eigenvalues in the light neutrino mass matrix, as
   in~\cite{Dreiner:2003yr,Chen:2008tc}.}  The coefficients
 multiplying each entry in eq.(\ref{eq:8}) can be in fact computed
 with the same technique introduced in~\cite{Duque:2008ah} for
 computing the down-quark and charged lepton masses.  In summary, the
 order of magnitude of the various entries in $M_\nu$ is determined by
 the appropriate powers of the small factor $\epsilon$ while, as we
 will see, the details of the mass spectrum are determined by
 non-hierarchical computable group theoretical coefficients, that only
 depend on the way the heavy FN states are assigned to $SU(5)$
 representations.

 \subsection{Coefficients of the Dirac and Majorana effective
   operators}

In this section we analyze the contributions of different effective
operators to $Y_\nu$ and to $M_R$, showing that a phenomenologically
acceptable structure, able to reproduce (approximately) the correct
mass ratios and to give reasonable neutrino mixing angles can be
obtained.

\begin{figure}[t!]
   \centering

\hspace{0.5cm}\includegraphics[scale=0.75]{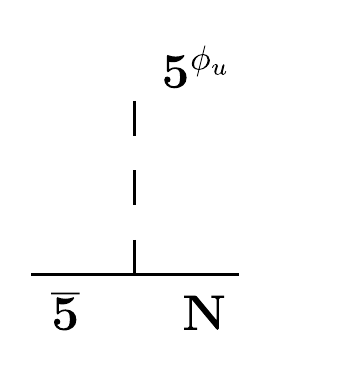} 
\hspace{2.3cm}\includegraphics[scale=0.75]{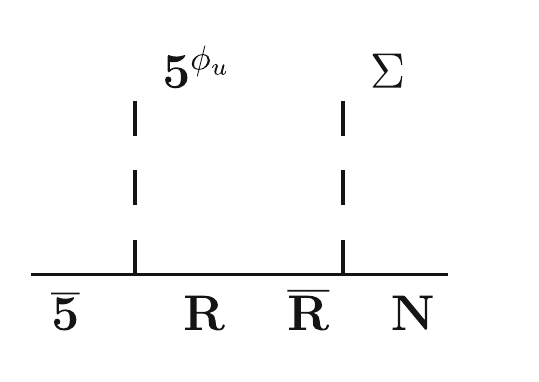}   
\hspace{1.7cm}\includegraphics[scale=0.75]{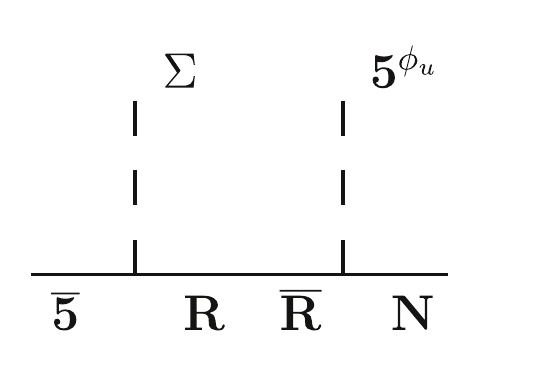} \\ [-5pt]
\leftline{\hspace{2.4cm}(a)\hspace{5.5cm}(b)\hspace{5.5cm}(c)} 

\vspace{10pt}

\includegraphics[scale=0.75]{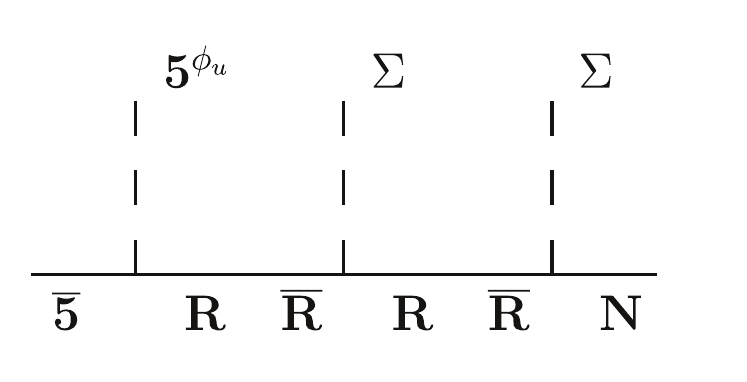}   
\includegraphics[scale=0.75]{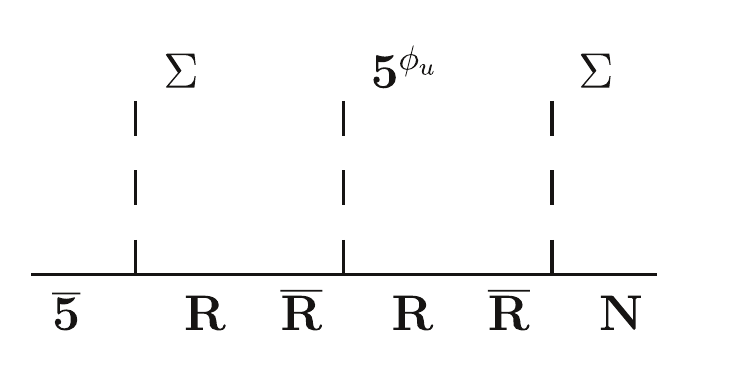} 
\includegraphics[scale=0.75]{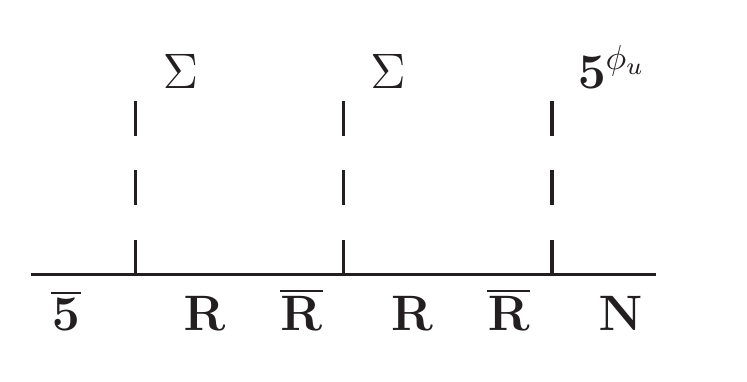} \\ [-5pt]
\leftline{\hspace{2.5cm} (d) \hspace{5.1cm} (e) \hspace{5.0cm} (f)}

\caption{ Diagrams contributing to $Y_\nu$ at different orders. The
  lowest order coefficient corresponding to diagram (a) is $y_0=3$.
  Diagrams (b)--(c) contribute at ${\cal O}(\epsilon^1)$ and yield the
  coefficients $y_1$ in the second column in
  Table~\ref{tab:1}. Diagrams (d)--(f) contribute at ${\cal
    O}(\epsilon^2)$ and give the coefficients $y_2$ in the fourth
  column of the table.}
  \label{fig:1}
\end{figure}

We assume that a large number of vectorlike FN fields exist in various
$SU(5)$ representations.  Since we assign the heavy Majorana neutrinos
to the adjoint $\mathbf{N}$, the possible FN field
representations $\mathbf{R}$ can be identified starting from the
following tensor products involving the representations of the
fields in the external lines (see the diagrams in Fig.~\ref{fig:1}):
\begin{align}
  \label{eq:9}
\mathbf{ \overline{5} \otimes {5_{\phi_u}} } =& \mathbf{
{1}\oplus{24} } \,, \\
\mathbf{\overline{5}} \>\> \mathbf{\otimes} \>\> \mathbf{\Sigma}\  =& \mathbf{
\overline{5}\oplus\overline{45}  \oplus \overline{70}}\,, \\
\mathbf{ {N} \> \otimes {\Sigma} } \>\> =& \mathbf{{1_S}}\oplus
\mathbf{{24_S}}  \oplus \mathbf{24_A} \oplus \mathbf{{75_S}}
  \oplus \mathbf{{126_A}}   \nonumber\\ 
&\oplus\mathbf{\overline{126_A}} \oplus \mathbf{{200_S}} \,,
\end{align}
where the subscripts $\scriptstyle\mathbf{S,\,A}$ in the last line
denote the symmetric or antisymmetric nature of the corresponding
representations.  We assume that all FN fields transform nontrivially
under $SU(5)$, and thus that no singlet exists and, for simplicity, we
restrict ourselves to representations with dimension less than 100,
which results in the following possibilities $\mathbf{R}=\mathbf{24}$,
$\mathbf{5}$, $\mathbf{45}$, $\mathbf{70}$.  

{\it Pointlike propagators:}\ Since the mass $M$ of
these fields is assumed to be larger than $\langle \Sigma_\pm\rangle
\sim \Lambda_{\text{GUT}}$, the contributions to the operators in
eq.~\eqref{eq:4} can be evaluated by means of insertions of effective
pointlike propagators. As in \cite{Duque:2008ah} we denote the
contractions of two vectorlike fields in the representation
$\mathbf{R}$, $\overline{\mathbf{R}}$ as
\begin{align}
  \left[\mathbf{R}_{d e\ldots}^{a b c\ldots}\overline{\mathbf{R}}_{l m n\ldots}^{p
        q\ldots}\right]
  =-\frac{i}{M}\mathcal{S}_{d e \ldots l m n\ldots}^{a b c\ldots p q \ldots}\,,
\end{align}
where all the indices are $SU(5)$ indices, and $\mathcal{S}$ is the appropriate
group index structure. The structures $\mathcal{S}$ for
$\left[\fv^a\,\bfv_b\right]$,
$\left[\for^{ab}_c\,\bfor^n_{lm}\right]$ and
$\left[\se^{ab}_c\,\bse^n_{lm}\right]$
(and for several other
$SU(5)$ representations) can be found  in Appendix A of
\cite{Duque:2008ah}. In addition we need the following
contractions
\begin{align}
  \label{eq:10S}
i M \left[  \mathbf{24}^a_b\,\mathbf{24}^l_m  \right]_{\mathbf{S}}&=
\left({\cal S}_{\mathbf{S}}\right)^{a\,l}_{b\,m}=
\frac{5}{2}\left[\delta^a_m\, \delta_b^l +
\delta^a_l\, \delta_b^m\right]
- \delta^a_b\, \delta_m^l \,,  \\
  \label{eq:10A}
i M\left[\mathbf{24}^a_b\,\mathbf{24}^l_m  \right]_{\mathbf{A}}&=
\left({\cal S}_{\mathbf{A}}\right)^{a\,l}_{b\,m}= \frac{5}{2}
\left[\delta^a_m\, \delta_b^l -\delta^a_l\, \delta_b^m\right]\,.
\end{align}
These two expressions are obtained by imposing the traceless condition
for the adjoint $\left({\cal S}_{\mathbf{S,A}}\right)^{a\,l}_{a\,m}
=\left({\cal S}_{\mathbf{S,A}}\right)^{a\,l}_{b\,l}=0$ and the
normalization factor is fixed by the requirement that the (subtracted)
singlet piece $\delta^a_b\, \delta_m^l$ in eq.~\eqref{eq:10S} provides
the proper singlet contraction, that is, by inserting the singlet in
the diagram of fig.\ref{fig:1}(b) we require that the operator
$\left(\mathbf{\overline{5}}_a \mathbf{5}_{\phi_u}^a\right)
\cdot\left(\mathbf{N}_l^j \mathbf{\Sigma}_j^l\right)$ is obtained with
unit coefficient.  

{\it Vertices:}\  All the vertices we need involve $\mathbf{5}_{\phi_u}$ or the
adjoint $\bSigma$ with the external fermions $\bfv$ and $\mathbf{N}$,
or with the FN representations $\mathbf{R}$ in the internal lines.
The vertices have the general form $-i \lambda {\cal V}$ where
$\lambda$ is universal for all vertices.  Including symmetry factors,
the relevant field contractions $ {\cal
  V}=\mathbf{R}\,\mathbf{5}_{\phi_u}\,\mathbf{R'} $ or ${\cal
  V}=\mathbf{R}\,\mathbf{\Sigma}\,\mathbf{R'}$, with
$\mathbf{R},\mathbf{R'}=\mathbf{5},\,\mathbf{24},\,\mathbf{45},\,\mathbf{70}$,
are:
\begin{eqnarray}
\label{eq:v5} &&
    \bfv_a \tf^a_b\fv^b \qquad  \!\!\!
\bfv_a \tf^c_b \for^{ba}_c \qquad  \!\!\!
\bfv_a \tf^c_b \se^{ba}_c 
\qquad \tf^a_c\tf^c_b
\left(\tf_{\mathbf{S,A}}\right)^b_a.
\\
\label{eq:v45} &&
\bfor_{ab}^c{\tf^\uparrow}^b_d\for^{da}_c \quad\    \quad
   \frac{1}{2}\, \bfor_{ab}^c{\tf^\downarrow}_c^d\for^{ba}_d  
 \\ \label{eq:v70} &&
\bse_{ab}^c{\tf^\uparrow}^b_d\se^{da}_c   \quad\   \quad
   \frac{1}{2}\, \bse_{ab}^c{\tf^\downarrow}_c^d\se^{ba}_d     
\quad\ \quad 
\bfor_{ab}^c\tf^b_d\se^{da}_c \,,  
\end{eqnarray}
where the vertices in the first line describe the couplings of the
external states ($\mathbf{\overline 5}$ and $\mathbf{N}$) with heavy
FN fields and flavons, while the last two lines involve only heavy FN
fields and flavons.  There are two inequivalent ways of contracting
the indices for the vertices involving the $\tf$ with pairs of $\for$
and $\se$ in the last two lines~\cite{Duque:2008ah}. They are distinguished in
eqs.~(\ref{eq:v45}) and (\ref{eq:v70}) by an up ($\tf^\uparrow$) or
down ($\tf^\downarrow$) arrow-label.  As explained
in~\cite{Duque:2008ah}, this can be traced back to the fact that these
representations are contained twice in their tensor products with the
adjoint.

\begin{figure}[h!!]
  \centering

\includegraphics[scale=0.85]{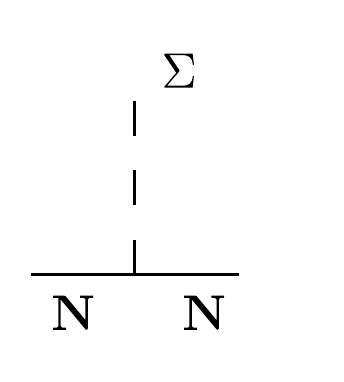} \hspace{1.0cm}
\includegraphics[scale=0.85]{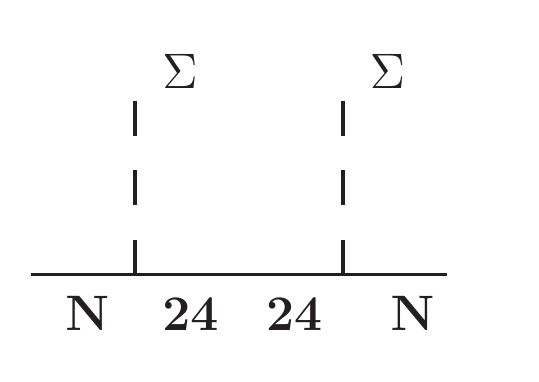} \hspace{1.0cm}
\includegraphics[scale=0.85]{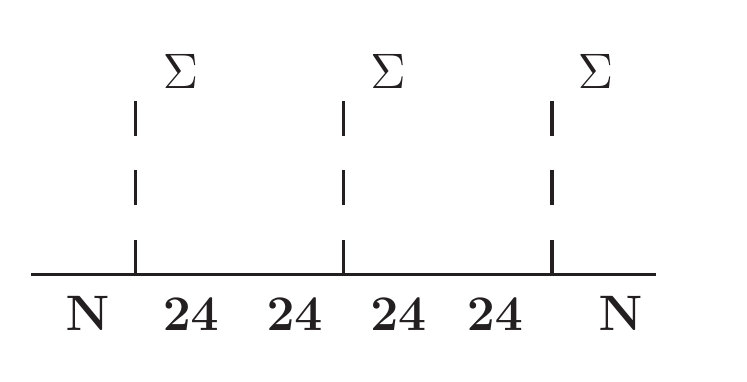} \\ [-5pt]

\leftline{\hspace{1.6cm}(a)\hspace{4.7cm}(b)\hspace{6.3cm}(c)}

\caption{
Diagrams contributing to  $M_R$ at different orders. The lowest order coefficient 
$r_1$ is obtained from diagram (a), $r_2$ from (b), and $r_3$ from (c).}
\label{fig:2}
\end{figure}

{\it Relevant multiplet components:}\ We write the $SU(5)\times
U(1)_F$ breaking vevs as
\begin{equation}
\label{eq:VEV}
\langle\mathbf{\Sigma_{\pm}}\rangle=
\frac{V}{\sqrt{60}}\times\operatorname{diag}(2,2,2,-3,-3)\,,
\end{equation}
where the factor $1/\sqrt{60}$ gives the usual normalization of the
$SU(5)$ generators,
$\operatorname{Tr}(\mathbf{R}^a\overline{\mathbf{R}^b})=(1/2)\delta^{ab}$,
and the coefficients of the left handed neutrino couplings to the
$SU(2)$ singlet $\nu\,\phi^0_u\,N_S$ and $SU(2)$ triplet
$\nu\,\phi^0_u\,N_T$ as well as the Majorana neutrinos mass terms
$N_{S,T}\,N_{S,T}$ are obtained by projecting the representations
$\mathbf{5}$, $\mathbf{5}_{\phi_u}$ and $\mathbf{N}$ onto the relevant
field components according to
\begin{eqnarray}
\label{eq:projections1}
\nu &=& - \overline{\mathbf{5}}_5 = - \delta_5^a \; \overline{\mathbf{5}}_a 
\\ 
\label{eq:projections2}
\phi^0_u &=&\mathbf{5}_{\phi_u}^5   =  \delta^5_b\; \mathbf{5}_{\phi_u}^b   
\\
\label{eq:projections3}
N_S &=& \frac{1}{\sqrt{60}}\;
\operatorname{diag}(2,\,2,\,2,\,-3,\,-3)\cdot \mathbf{N}_{24}\,.
\\
N_T &=& \frac{1}{\sqrt{60}}\;
\operatorname{diag}(0,\,0,\,0,\,\sqrt{15},\,-\sqrt{15})
\cdot \mathbf{N}_{3}\,.
\label{eq:projections4}
\end{eqnarray}
where the subscripts in $\mathbf{N}_{24}$ (singlet) and
$\mathbf{N}_{3}$ (neutral component of the triplet) refer to the
corresponding $SU(5)$ generators.  The assumption of a unique heavy
mass parameter $M$ for the FN fields and of universality of the
fundamental scalar-fermion couplings $\lambda$ yield a remarkable
level of predictivity.  In particular, for the vertices involving
$\mathbf{\Sigma_\pm}$ we can always reabsorb $\lambda V \to V$. This
leaves just an overall power of $\lambda$ common to all effective
Yukawa operators that involve one insertion of the Higgs multiplet
$\mathbf{5}_{\phi_u}$ (see the diagrams in Figs.~\ref{fig:1}) and no
$\lambda$ at all for the contributions to $M_R$, (see the diagrams in
Figs.~\ref{fig:2}).

The contributions to $Y_\nu$ and $M_R$ at different orders can be
computed using the vertices $\mathcal{V}$ given in
eqs.~\eqref{eq:v5}-\eqref{eq:v70} and the relevant group structures
$\mathcal{S}$ in eqs.~\eqref{eq:10S}, \eqref{eq:10A} and in Appendix A
of~\cite{Duque:2008ah}, that account for integrating out the heavy FN
fields. Additionally, the multiplets $\mathbf{\bar 5}$, $\mathbf{N}$,
and $\mathbf{5}_{\phi_u}$ in the external legs of the diagrams must be
projected on the relevant components according to
eqs.~\eqref{eq:projections1}-\eqref{eq:projections4} and the flavons
$\mathbf{\Sigma_\pm}$ have to be projected onto the vacuum according to
eq.~\eqref{eq:VEV}.

We have evaluated the $Y_\nu$ including the contributions up to
$\mathcal{O}(\epsilon^2)$ that are diagrammatically depicted in
Figs.~\ref{fig:1}: \ref{fig:1}(a) $\mathcal{O}(\epsilon^0)$;
\ref{fig:1}(b)--\ref{fig:1}(c) $\mathcal{O}(\epsilon^1)$;
\ref{fig:1}(d)--\ref{fig:1}(f) $\mathcal{O}(\epsilon^2)$.  $M_R$ has
been computed including contributions with three insertions 
of the flavons $\Sigma_\pm$ 
corresponding to the diagrams in Figs.~\ref{fig:2}: \ref{fig:2}(a)
$\mathcal{O}(\epsilon)$; \ref{fig:2}(b) $\mathcal{O}(\epsilon^2)$;
\ref{fig:2}(c) $\mathcal{O}(\epsilon^3)$.  At each specific order, 
the contributions to specific entries in $Y_\nu$ and $M_R$ 
can be written as 
\begin{eqnarray}
  \label{eq:orders}
  Y_\nu^{(i)} &=&  \lambda\,\alpha^{i+1}\,\epsilon^i \cdot  
\left(y^S_i +y^T_i\right)\,,
\\
M_R^{(i)}& =& V\,\alpha^{i+3}\,\epsilon^i 
\cdot \left(r^S_{i+1}+r^T_{i+1} \right)= 
M\,\alpha^{i+3}\,\epsilon^{i+1}
\cdot \left(r^S_{i+1}+r^T_{i+1} \right)\,,
 \end{eqnarray}
 where $\alpha= 1/\sqrt{60}$ is the normalization factor for
 $\mathbf{\Sigma}$ and for the $N_{S,T}$ in the adjoint, $V=M\epsilon$
 with $V$ defined in eq.~(\ref{eq:VEV}), and $y_i^{S,T}$ and
 $r_{i+1}^{S,T}$ are the nontrivial group theoretical coefficients,
 that we have computed for $i=0,1,2$ and for the singlet $(S)$ and triplet
$(T)$ contributions to the seesaw Lagrangian. 
The corresponding
 results for $y_i^{S,T}$ are given in Table~\ref{tab:1} (where we have
 followed the notation of~\cite{Duque:2008ah}), while the results for
 $r_{i+1}^{S,T}$ are given in Table~\ref{tab:2}.

%

\renewcommand{\arraystretch}{1.28}
\begin{table}[t!]
  \centering
  \begin{tabular}{c|c|c|c|c|c|c|c|c|c}
    $\mathbf{\bar 5}_1$ & $\mathbf{\bar 5}_2$ & $\mathbf{\bar 5}_3$ 
& $\mathbf{10}_1$ & $\mathbf{10}_2$ & $\mathbf{10}_3$ &
    $\mathbf{5}_{\phi_u}=-\mathbf{5}_{\phi_d}$&$N_1$&$N_2$ &$N_3$\\\hline
    $-\frac{29}{10}$& -$\frac{9}{10}$ & $\frac{31}{10}$&
$\frac{13}{10}$&$\frac{3}{10}$&$-\frac{7}{10}$&$\frac{7}{5}$&
$\frac{5}{2}$&$-\frac{1}{2}$&$-\frac{11}{2}$\\
  \end{tabular}
  \caption{$F$--charges obtained with $n=-6$ in eq.~\eqref{eq:2}, 
and $m_1=2$, $m_2=-1$, and $m_3=-6$ in eq.~\eqref{eq:3}.} 
  \label{tab:3}
\end{table}
\renewcommand{\arraystretch}{1}

We have searched for all possible charge assignments with absolute
values of the $F$ charges smaller than 10, and we have examined the
resulting neutrino mass matrices.  We have found some promising
possibilities.  If we choose, for example, in eqs.~(\ref{eq:5}) and
(\ref{eq:6}), $n=-6$ and $m_1=2$, $m_2=-1$, $m_3=-6$, we obtain the
$F$--charges shown in Table~\ref{tab:3}, which can be obtained from
the set given in eq.~\eqref{eq:1} through the redefinitions
eqs.~\eqref{eq:shift} with $a_{-6}=21/10$.

According to eq.~\eqref{eq:5}, this set of $F$--charges 
gives the following orders of magnitude for $Y_\nu$:
\begin{align}
  \label{eq:12}
    Y_{\nu}\sim&\lambda
  \begin{pmatrix}
\epsilon   & \epsilon^2    & \epsilon^7 \\
\epsilon^3 & 1           & \epsilon^5 \\
\epsilon^7 &  \epsilon^4 & \epsilon \\
  \end{pmatrix}.
\end{align}
Neglecting terms of $\mathcal{O}(\epsilon^4)$ and higher, 
including the coefficients $y_i^{S,T}$ and the 
appropriate powers of the normalization factor $\alpha$, this reads:
\begin{align}
 Y_\nu^{S,T}\,\approx\,&\lambda\, \alpha\,  \begin{pmatrix}
y_1(\alpha\epsilon)  & y_2(\alpha\epsilon)^2 & 0 \\
y_3(\alpha\epsilon)^3 & y_0  & 0 \\
0 & 0  & y_1(\alpha\epsilon) \\
  \end{pmatrix}^{S,T}.
\end{align}
where the superscript $^{S,T}$ outside the matrix is a shorthand for
$y_i^{S,T}$ inside the matrix.  Similarly, according to
eq.~\eqref{eq:6} and \eqref{eq:8} we have for the entries in $M_R$ the
following orders of magnitude:
\begin{align}
  \label{eq:13}
M_R\,\sim\, & 
V\,   
  \begin{pmatrix}
\epsilon^4 & \epsilon^1 & \epsilon^2\\
\epsilon^1 & \epsilon^0 &\epsilon^5   \\
\epsilon^2 & \epsilon^5 &\epsilon^{10}   \\
  \end{pmatrix}.
\end{align}
Neglecting terms of $\mathcal{O}(\epsilon^4)$ and higher, and 
taking into account the coefficients $r_i^{S,T}$ and $\alpha$, we obtain
\begin{align}
\label{eq:14}
 M_R^{S,T} \approx\, & \, V \, \alpha^{3}\, 
\begin{pmatrix}
0   & r_2 (\alpha\epsilon) & r_3(\alpha\epsilon)^2\\
r_2(\alpha\epsilon) & r_1 &0   \\
r_3(\alpha\epsilon)^2 & 0 &0   \\
  \end{pmatrix}^{S,T} \,.
\end{align}
According eq.~\eqref{eq:7}, the resulting light neutrino mass matrix  then is 
\begin{align}
\label{eq:Mnuappx}
M_\nu\approx& - \frac{v^2\sin^2\beta}{\alpha V}
  \,\lambda^{2}\sum_{S,T} \left[ \frac{1}{r_1\,r_3}\, \begin{pmatrix}
0    &  0 & y_1^2r_1 \\
0    & \phantom{\Big|}\!\!y_0^2r_3 & -y_0y_1r_2 \\ 
y_1^2 r_1 & -y_0y_1r_2 & \frac{1}{r_3}\, y_1^2 r_2^2
  \end{pmatrix}\right],
\end{align}
where we have neglected in each entry corrections of ${\cal
  O}(\alpha\epsilon)^2$ and higher, and we have suppressed the
subscripts $^{S,T}$ not to clutter the expression. It is remarkable
that at leading order the structure of the light neutrino mass matrix
remains determined only in terms of the group theoretical coefficients
$y_i^{S,T}$ and $r_i^{S,T}$, and in particular it does not depend on the
hierarchical parameter $\epsilon$. Let us also note that this matrix
corresponds to the two zero--texture type of neutrino mass matrix
discussed in~\cite{Mohanta:2006xd}.  As regards the scale $\alpha V$
appearing in the denominator of eq.~\eqref{eq:Mnuappx}, it can be
directly related with the unification scale, defined as the mass scale
of the leptoquarks gauge fields $M_X=M_Y$~\cite{Bailin:1986wt}:
\begin{align}
\label{eq:Lgut}
  \Lambda_{GUT}=M_X
=5\,g_5\,\alpha\, V\,,
\end{align}
where $g_5\approx 0.7$ is the unified gauge coupling at
$\Lambda_{GUT}\simeq10^{16}$.


It is remarkable to note that both $Y_\nu$ and $M_R$ are hierarchical,
with the first one having a hierarchy between its eigenvalues of
${\cal O}(\alpha\epsilon)$ and the second one of ${\cal
  O}(\alpha\epsilon^2)$.  The light neutrino mass matrix computed
naively (and erroneously, see Section~\ref{sec:signs}) from the
effective seesaw operator using only the charges of the
$\overline{\mathbf{5}}_I$ multiplets, would also be hierarchical.
However, the resulting $M_\nu$ is not hierarchical, and in fact at
leading order it does not depend at all on $\epsilon$ but only on the
group theoretical coefficients $y_i^{S,T}$ and $r_i^{S,T}$.  It is precisely the
presence of $F$ charges of both signs for the fields and for the two
flavons that yields the possibility of obtaining non-hierarchical
neutrino masses and large mixing angles, although the whole scenario
is defined at the fundamental level in terms of a small hierarchical
parameter $\epsilon$.

Let us comment at this point that, as it is discussed
in~\cite{Duque:2008ah}, corrections from sets of higher order diagrams
to the various entries in $Y_\nu$ and $M_R$ can generically be quite
sizable, although suppressed by higher powers of $\epsilon$. This is
because at higher orders the number of diagrams contributing to the
various operators proliferate, and the individual group theoretical
coefficients also become generically much larger, as can be seen in
tables~\ref{tab:1} and~\ref{tab:2}.  By direct evaluation of higher
orders corrections, the related effects were estimated
in~\cite{Duque:2008ah} to be typically of a relative order $\sim 20\%
- 30\%$. To take into account the possible effects of these
corrections, we allow for a $\sim 25\% $ uncertainty in  the 
final numerical results.

\section{Numerical analysis}
\label{sec:numerical}


\begin{table}[t!]
  \centering
\begin{tabular}{|l|r|r||l|r|r|}\hline
\phantom{$\Big|$}$\epsilon^1$ & $y_1^S$& $y_1^T$&\ $\epsilon^2$ &$y_2^S$&$y_2^T$ \  \\
\hline\phantom{$\Big|$}
$\!\![\mathbf{5}^{\phi_{u}}\Sigma]$& & &$[\mathbf{5}^{\phi_{u}}\Sigma\Sigma]$&$ $&$$ \ \\  [3pt]
$O(\epsilon;\mathbf{24_S})$&$-15$&$-15\sqrt{15}$&$O(\epsilon^{2};\mathbf{24_S},\mathbf{24_S})$&$-75$&$-225\sqrt{15}$\\[2pt]
$O(\epsilon;\mathbf{24_A})$&$\ 0$&$0$&$O(\epsilon^{2};\mathbf{24_A},\mathbf{24_S})$& $0$&$0$\\ [2pt]
&&&$O(\epsilon^{2};\mathbf{24_S},\mathbf{24_A})$& $0$&$0$\\ [2pt]
&&&$O(\epsilon^{2};\mathbf{24_A},\mathbf{24_A})$& $0$&$0$\\  [3pt]
\hline \phantom{$\Big|$}
$[\Sigma\mathbf{5}^{\phi_{u}}]$& & &$[\Sigma\mathbf{5}^{\phi_{u}}\Sigma]$&$ $&$$ \\ [3pt]
$O(\epsilon;\mathbf{5})$&$\ -9$&$-3\sqrt{15}$&$O(\epsilon^{2};\mathbf{5},\mathbf{24_S})$&$-45$&$-45\sqrt{15}$\\  [2pt]
&&&$O(\epsilon^{2};\mathbf{5},\mathbf{24_A})$& $\ \ 0$&$0$\\  [2pt]
$O(\epsilon;\mathbf{45})$&$\  75$&$-15\sqrt{15}$&$O(\epsilon^{2};\mathbf{45},\mathbf{24_S})$&$300$&$-180\sqrt{15}$\\ [2pt]
&&&$O(\epsilon^{2};\mathbf{45},\mathbf{24_A})$&$\ \ 0$&$0$\\ [2pt]
$O(\epsilon;\mathbf{70})$&$-225$&$-15\sqrt{15}$&$O(\epsilon^{2};\mathbf{70},\mathbf{24_S})$&$-900$&$-180\sqrt{15}$\\ [2pt]
&&&$O(\epsilon^{2};\mathbf{70},\mathbf{24_A})$& $\ \ 0$&$0$\\ [2pt]
\cline{4-6}& &\phantom{$\Big|$}
&$[\Sigma\Sigma\mathbf{5}^{\phi_{u}}]$&$ $&$$                    \\ [3pt]
&&&$O(\epsilon^{2};\mathbf{5},\mathbf{5})$& $-27$&$-9\sqrt{15}$             \\ [2pt]
&&&$O(\epsilon^{2};\mathbf{5},\mathbf{45})$& $225$ &$-45\sqrt{15}$            \\ [2pt]
&&&$O(\epsilon^{2};\mathbf{5},\mathbf{70})$& $-675$&$-45\sqrt{15}$           \\ [2pt]
&&&$O(\epsilon^{2};\mathbf{45},\mathbf{5})$& $225$&$75\sqrt{15}$             \\ [2pt]
&&&$O^{\uparrow}(\epsilon^{2};\mathbf{45},\mathbf{45})$& $1425$&$-285\sqrt{15}$ \\ [2pt]
&&&$O^{\downarrow}(\epsilon^{2};\mathbf{45},\mathbf{45})$&$525$&$-105\sqrt{15}$\\ [2pt]
&&&$O(\epsilon^{2};\mathbf{45},\mathbf{70})$& $1125$&$75\sqrt{15}$           \\ [2pt]
&&&$O(\epsilon^{2};\mathbf{70},\mathbf{5})$& $-675$&$-225\sqrt{15}$           \\ [2pt]
&&&$O(\epsilon^{2};\mathbf{70},\mathbf{45})$& $-1125$&$225\sqrt{15}$          \\ [2pt]
&&&$O^\uparrow(\epsilon^{2};\mathbf{70},\mathbf{70})$& $-4725$&$-315\sqrt{15}$ \\ [2pt]
&&&$O^\downarrow(\epsilon^{2};\mathbf{70},\mathbf{70})$& $-675$&$-45\sqrt{15}$\\ [3pt]
\hline\phantom{$\Big|$}
$\Sigma_{\mathbf{R}}O(\epsilon; \mathbf{R})$&$-174$&$-48\sqrt{15}$&$\Sigma_{\mathbf{R}} O(\epsilon^{2}; \mathbf{R})$&$-5097$&$-1329\sqrt{15}$\\ [3pt]
\hline
\end{tabular}
\caption{Operators contributing to  ${Y}_\nu=\sum_i {Y}^{(i)}_\nu$ 
  at ${\cal O}(\epsilon)$ and  ${\cal O}(\epsilon^2)$ and  values of 
  the corresponding coefficients 
  $y_i =Y^{(i)}_\nu/\left(\lambda\,\alpha^{i+1}\, \epsilon^i\right)$
  for the singlet $(S)$ and triplet $(T)$ components.  
  The value of the  ${\cal O}(1)$ coefficients are 
  $y_0^S=3$ and $y_0^T=\sqrt{15}$.
}
\label{tab:1}
\end{table}
%

%
\begin{table}[t!] 
  \centering
\begin{tabular}{|l|c||l|c||l|c|}\hline
\phantom{$\Big|$}$\epsilon^1$&$(r_1^S,r_1^T)$
& $\epsilon^2$ & $(r_2^S,r_2^T)$
& $\epsilon^3$ & $(r_3^S,r_3^T)$
\\\hline
\phantom{$\Big|$}
$\!\![\Sigma]$&&$[\Sigma\Sigma]$&& $[\Sigma\Sigma\Sigma]$&$$ \\  [6pt]
\phantom{$\Big|$}$O(\epsilon;\mathbf{ 24})$ & $(-30,-90)$
&$O(\epsilon^{2};\mathbf{ 24_S})$& $(-150,-1350)$
&$O(\epsilon^{3};\mathbf{24_S},\mathbf{24_S})$& $(-750,-20250)$ \\  [4pt]
&&$O(\epsilon^{2};\mathbf{ 24_A})$& $(0,0)$
&$O(\epsilon^{3};\mathbf{24_A},\mathbf{24_S})$& $(0,0)$\\ [4pt]
&& &&$O(\epsilon^{3};\mathbf{24_S},\mathbf{24_A})$& $(0,0)$\\ [4pt]
&& &&$O(\epsilon^{3};\mathbf{24_A},\mathbf{24_A})$& $(0,0)$\\ [4pt]
\hline \phantom{$\Big|$}
$\Sigma_{\mathbf{R}} O(\epsilon; \mathbf{R})$&$(-30,-90)$
&$\Sigma_{\mathbf{R}} O(\epsilon^{2}; \mathbf{R})$&$(-150,-1350)$
&$\Sigma_{\mathbf{R}} O(\epsilon^{3}; \mathbf{R})$&$(-750,-20250)$ \\ \hline
\end{tabular}
\caption{
  Operators contributing to  ${M}_R=\sum_i M^{i}_R$  at 
  ${\cal O}(\epsilon)$, ${\cal O}(\epsilon^2)$  
  and  ${\cal O}(\epsilon^3)$, and values of the corresponding coefficients 
  $r_i =M_R^{(i-1)}/\left(V\alpha^{i+2}\,\epsilon^{i-1}\right)$ for the singlet $(S)$ and triplet $(T)$ components.  
}
\label{tab:2}
\end{table}

Allowing for all the contributions listed in Table~\ref{tab:1}, the
resulting coefficient at ${\cal O}(\epsilon)$ for $Y_\nu^S$ would be
$y_1^S=-174$ that is too large to reproduce the neutrino oscillation
data.  We will then assume that only some contributions are present.
This is easily achieved by assuming that no FN fields exist in the
representations $\mathbf{70}_{39/10}$, $\mathbf{70}_{-41/10}$,
$\mathbf{45}_{39/10}$ and $\mathbf{45}_{-41/10}$, and this results in
much smaller coefficients $y_1^S=-24$ and $y_1^T=-18\sqrt{15}$ that
are determined by the $y_1^{S,T}$ entries in the first and third lines
in Table~\ref{tab:1}, and that are the one we will use henceforth.
(The absence of these representations also implies that several
contributions to the higher order coefficient $y_2^{S,T}$ are absent,
which yields much smaller values $y_2^{S,T}\sim 10^2$ instead than
$\sim 10^3$, see Table~\ref{tab:1}.  In any case, since at leading
order $M_\nu$ eq.~(\ref{eq:Mnuappx}) does not depend on $y_2^{S,T}$,
this only affects the higher order corrections.) As regards the
contributions to $M_R$, they arise only from insertions of the
$\mathbf{24}$, and thus they are not affected by the absence of
$\mathbf{70}$ and $\mathbf{45}$.

By using in
eq.~(\ref{eq:Mnuappx}) $(y_0^S,y_0^T)=(3,\sqrt{15})$, 
$(y_1^S,y_1^T)=(-24,-18\sqrt{15})$ 
and the values of $(r_i^S,r_i^T)$ given  in
table~\ref{tab:2}, we  obtain
\begin{align}
\label{eq:15}
 M_\nu \;\approx\; & - 5\, 
\left(\lambda\,\sin\beta\right)^2
\frac{g_5\, v^2}{\Lambda_{\text{GUT}}}
   \, 
\begin{pmatrix}
0    & 0 & -1.0 \\
0    & -0.47 & -0.68 \\
-1.0 & -0.68 & -1.0
  \end{pmatrix}.
\end{align}
With $v=175\,$GeV and $\Lambda_{\text{GUT}}\approx 10^{16}\,$GeV
 the numerical value of the
prefactor is $ 
\approx 0.008\,(\sin\beta\,\lambda)^2\,$eV.  For $\tan\beta\approx10$
($\tan\beta\approx1$) the atmospheric mass scale $\approx 0.05\,$eV
can then be reproduced for acceptable values of the coupling
$\lambda\sim 1.9\;(2.7)\,$.

Our model is based on the successful model for the $d$-quark and
leptons masses discussed in Ref.~\cite{Duque:2008ah}, and we have
checked that the absence of the representations that we have forbidden
here do not affect the results of this previous study.  In particular,
by using the coefficients calculated in Ref.~\cite{Duque:2008ah} we
have for the matrix of the charged leptons Yukawa couplings
\begin{align}
  Y^e\simeq
  \begin{pmatrix}
  \epsilon^4 & \epsilon^5 & \epsilon^4 \\
 -2.9\epsilon^3 & 3.8\epsilon^2 
& 10.2\epsilon^3 \\
 -7.6\epsilon^3 & 9.2\epsilon^2 & 2.3\epsilon 
  \end{pmatrix}\,.
\end{align}
To compute neutrino mixing matrix
$U_{PMNS}=U_\nu\left(V_L^e\right)^\dagger$, besides the matrix $U_\nu$
that diagonalizes $M_\nu$ in eq.~\eqref{eq:15}, we also need $V_L^e$
that diagonalizes the left-handed product ${Y^e}{Y^e}^\dagger$. We
obtain
\begin{equation}
  \label{eq:Ue}
V_L^e \sim 
  \begin{pmatrix}
1.     & 10^{-5} &10^{-5}  \\
10^{-5} & -1 & 0.02 \\
 10^{-5} & 0.02 & 1
  \end{pmatrix}.
\end{equation}
that is approximately diagonal, and thus $U_{PMNS}\approx U_\nu$.
Allowing for a $\sim 25\%$ numerical uncertainty in the entries of the
matrix in eq.~\eqref{eq:15}, we find that it is possible to fit the
neutrino oscillation data, with the exception of
$\sin^2\theta_{12}\sim 0.5$ for which a particularly large corrections
is needed. Finally, the mass of the lightest heavy singlet and triplet
neutrino states can be obtained from eq.~\eqref{eq:14} and are
\begin{align}
   & M_1^S\ \approx\ 5\times 10^{11}\, \text{GeV}\,,\\ 
&  M_1^T\approx 1.5\times 10^{13}\, \text{GeV}\,.
\end{align}
In particular the mass of the singlet Majorana neutrino is of the
right order of magnitude to allow for thermal
leptogenesis~\cite{Davidson:2008bu}.


\section{Conclusions}

We have extended the $SU(5)\times U(1)_F$ model for charged fermion
masses studied in Ref.~\cite{Duque:2008ah} to include neutrino masses.
This has been done by means of an appropriate redefinition of the
$U(1)_F$ charges that, while it leaves unchanged the Yukawa matrices
for the charged fermions, it also forbids at all orders $\Delta B\neq
0$ and $\Delta L=1$ operators, while allowing for $\Delta L=2$
Majorana mass terms. Thus, $R$-parity is enforced as an exact
symmetry, but at the same time the seesaw mechanism can be embedded
within the model.  Our construction is severely constrained by two
theoretical requirements. First, the $SU(5)$ GUT implies that the $F$
charges of the lepton doublets and $d$-quarks singlets, as well as the
$F$ charges of the quark-doublets and lepton singlets are the same,
reducing drastically the freedom one has in the SM. Second, we have
assumed universality of all the fundamental scalar-fermion couplings,
which basically implies that the model has only one free parameter,
that is the ratio between the $U(1)_F$ breaking vevs and the messenger
scale $M$. In spite of these serious restrictions, we have shown that
by assigning the $U(1)_F$ breaking flavons to the adjoint of $SU(5)$,
computable group theoretical coefficients arise that, at leading
order, determine the structure of the neutrino mass matrix in a
parameter independent way. This structure yields a reasonable first
approximation to the measured neutrino parameters.  However, higher
order corrections can be large, and should be taken into account for a
more precise quantitative comparison with observations.  In our model,
hierarchical heavy Majorana neutrinos naturally coexist with
non-hierarchical light neutrinos, the atmospheric scale is easily
reproduced for natural values of the parameters, and the mass of the
lightest heavy neutral states, that lies about five order of
magnitude below the GUT scale, is optimal for leptogenesis.  

At the quantitative level, the predictivity of the model clearly 
relies on the assumption of universality of the Yukawa couplings. We
have not put forth any speculation concerning the fundamental physics
that might underlie such a strong assumption, but have merely adopted
it as a working hypothesis to highlight how a theory of calculable
`order one coefficients' might actually emerge in GUT models relying
just on a generalized FN mechanism.  Needless to say, by relaxing the
assumption of universality by a certain quantitative amount, all the
predictions would acquire a correspondent numerical uncertainty,
although the main qualitative features of the model will remain
unchanged.

\section{Acknowledgments}
We are grateful to W. Tangarife for his participation in the first
stages of this work.  This research has been supported in part by
Sostenibilidad-UdeA/2009 grants: IN10140-CE, IN10157-CE.

\bibliographystyle{h-physrev4}
\bibliography{su5}

\end{document}